\documentstyle[aps,prb,amsfonts,epsfig,preprint]{revtex}
\begin{document}

\draft
\preprint{MSC-97-07}

\title{Capacitively coupled Josephson-junction chains: straight and
slanted coupling}
\author{Mahn-Soo Choi}
\address{Department of Physics and Pohang Superconductivity Center,
Pohang University of Science and Technology, Pohang 790-784, Korea}
\maketitle

\begin{abstract}
Two chains of ultrasmall Josephson junctions, coupled
capacitively with each other in the two different ways, straight
and slanted coupling, are considered.
As the coupling capacitance increases, regardless of the coupling
scheme, the transport of particle-hole pairs in the system is
found to drive the quantum-phase transition at zero temperature, which
is a insulator-to-superfluid transition of the pairs
and belongs to the Berezinskii-Kosterlitz-Thouless universal class.
The different underlying transport mechanisms for the two coupling
schemes are reflected in the difference between the transition points.
\end{abstract}

\pacs{PACS Numbers: 74.50.+r, 67.40.Db, 73.23.Hk}

\newcommand\CC{{\Bbb{C}}}
\newcommand\UU{{\Bbb{U}}}
\newcommand\varS{{\cal S}}
\newcommand\varG{{\cal G}}
\newcommand\bfm{{\bf m}}
\newcommand\bfq{{\bf q}}
\newcommand\bfr{{\bf r}}
\newcommand\bfJ{{\bf J}}
\newcommand\avgl{\left\langle}
\newcommand\avgr{\right\rangle}
\newcommand\hate{\hat{e}}
\newcommand\hatU{\widehat{U}}
\newcommand\tilU{\widetilde{U}}
\newcommand\tilC{\widetilde{C}}


\section{Introduction}\label{sec:intro}

\begin{figure}
\begin{center}
\epsfig{file=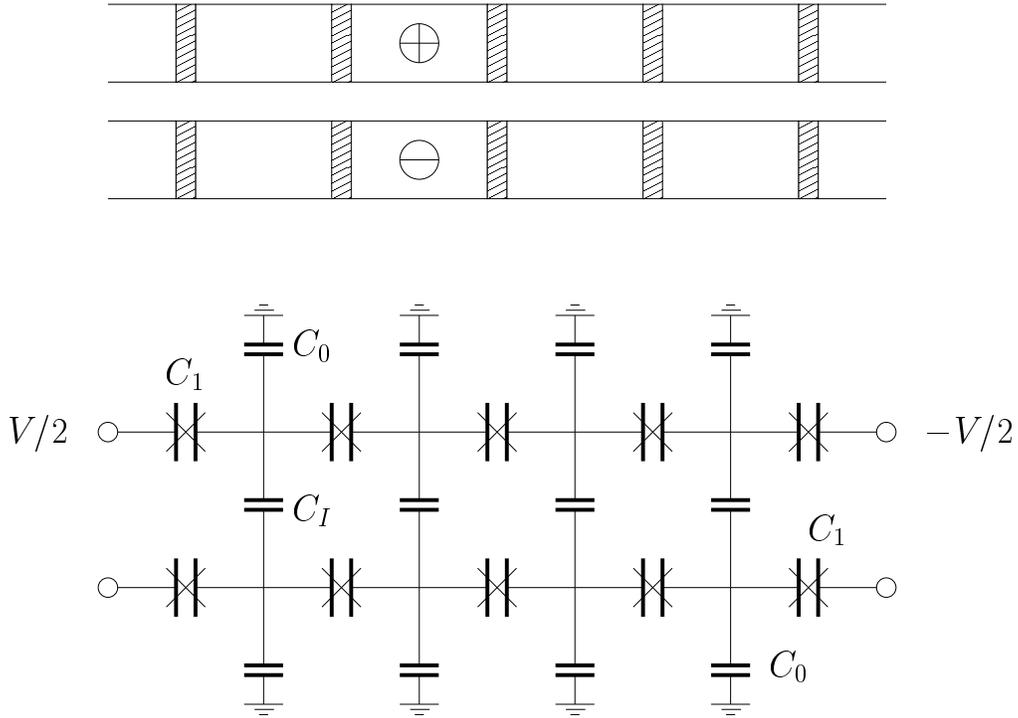,clip=,width=0.9\columnwidth}
\end{center}
\caption{Schematic picture and equivalent circuit of the system with
straight inter-chain coupling.  After Averin, Korotkov, and Nazarov,
Phys. Rev. Lett. {\bf 66}, 2818  (1991).}
\label{fig:1}
\end{figure}

Systems of ultrasmall tunnel junctions composed of metallic or
superconducting electrodes have attracted considerable interest owing
to the significant roles of the Coulomb interaction in them.
First of all, a sufficiently large charging energy leads to the
Coulomb blockade effect which exhibits single charge
(electron or Cooper pair) tunneling~\cite{Graber92}.  
In order for this tunneling to occur, however, it should be
energetically favorable with respect to the electrostatic energy of
the system.  Otherwise, more complex elementary processes that involve
several charge-tunneling events become dominant.
In particular, a recent theoretical prediction~\cite{Averin91a} and
an experimental demonstration~\cite{Matter97} have revealed the {\em
cotunneling} of electron-hole pairs in two one-dimensional (1D)
metallic tunnel-junction arrays coupled by large inter-array
capacitances (see Fig.~\ref{fig:1}).
Such cotunneling of electron-hole pairs results in the remarkable
effect of current drag: 
The current fed through either of the chains induces a secondary
current in the other chain.
The primary and the secondary currents are comparable in magnitude, but
opposite in direction.

A similar current drag effect has also been observed in a slightly
different configuration of the system, in which each electrode in one
array is coupled aslant to two adjacent electrodes in the other array
(slanted coupling; see Fig.~\ref{fig:2})~\cite{Delsin96}.
Unlike the case mentioned above (straight coupling; see
Fig.~\ref{fig:1}), where the low-energy state can be preserved only if
the electron and the hole tunnel simultaneously (cotunneling), in this
case the low-energy state can be preserved for sequential tunneling of
the electron and the hole. 
It has been suggested that this correlated sequential tunneling might
be more likely than the second-order process of cotunneling via a
quantum-mechanical virtual state.

More interestingly, when the tunneling junctions are composed of
ultrasmall superconducting grains, the counter part of the
electron-hole pair becomes the pair of an excess and a deficit Cooper
pair, which will be simply called a particle-hole pair. 
Furthermore, in such ultrasmall Josephson-junction systems, the
competition between the charging energy and the Josephson coupling
energy is well known to bring about the noble effects of quantum
fluctuations~\cite{Bradle84,Faziox91,BJKim95,mschoi98a}.
In a previous work~\cite{mschoi97a}, it was proposed that,
combined with these quantum-fluctuation effects, the cotunneling of
particle-hole pairs in capacitively coupled 1D Josephson junction
arrays (JJAs) drives the insulator-to-superfluid
transition of the pairs~\cite{mschoi97a}.

In this paper, the previous work~\cite{mschoi97a} on two
capacitively coupled 1D JJAs is extended to consider both straight
and slanted couplings.   The focus will be on the similarities and
differences between the two coupling schemes.
It is shown that on long-time and
long-length scales, the two cases are indistinguishable; as the coupling
capacitance increases, both of them exhibits
Berezinskii-Kosterlitz-Thouless (BKT)-type insulator-to-superfluid
transition, whose superfluid phase is uniquely characterized by the
condensation of particle-hole pairs and by perfect drag of
supercurrents along the two chains.
The correlated tunneling nature of the elementary process in the
slanted case is reflected by its lower transition point.

Capacitively coupled JJAs can presumably be realized in experiment,
even by current techniques.  Recent advances in microfabrication
techniques have already made it possible to create large arrays of
ultrasmall Josephson junctions~\cite{Schonx90}.
Furthermore, the experimental realization of the capacitively coupled
submicron metal-junction arrays~\cite{Matter97,Delsin96} illustrates
that large interarray capacitances can
also be fabricated between two JJAs.
In passing, capacitive coupling should be distinguished from
Josephson coupling (allowing interarray Cooper-pair tunneling).
The quantum-fluctuation effects in the latter case, as in the case of
quantum Josephson ladders, have been studied in the
literature~\cite{Granat90}.

The paper is organized as follows:
In Section \ref{sec:model}, the models for the systems with straight
or slanted couplings and the regions of interest in parameter space
are defined.
Section \ref{sec:2DVG} is devoted to the transformation of the models
into equivalent 2D systems of classical vortices.  In Section
\ref{sec:sigma}, the conductivity of the system is examined in the
vortex representation.
The results from Sections \ref{sec:2DVG} and \ref{sec:sigma} provide
the bases of this work on which the phase transition and the
current-drag effect are discussed in Section \ref{sec:QPT}.
Section \ref{sec:QPT} constitutes the main part of this paper and
present a thorough discussion of the role of particle-hole pairs in
the quantum-phase transition and the transport in the system.
Finally, the paper is concluded in Section \ref{sec:concl}.

\section{Models}\label{sec:model}

\begin{figure}
\begin{center}
\epsfig{file=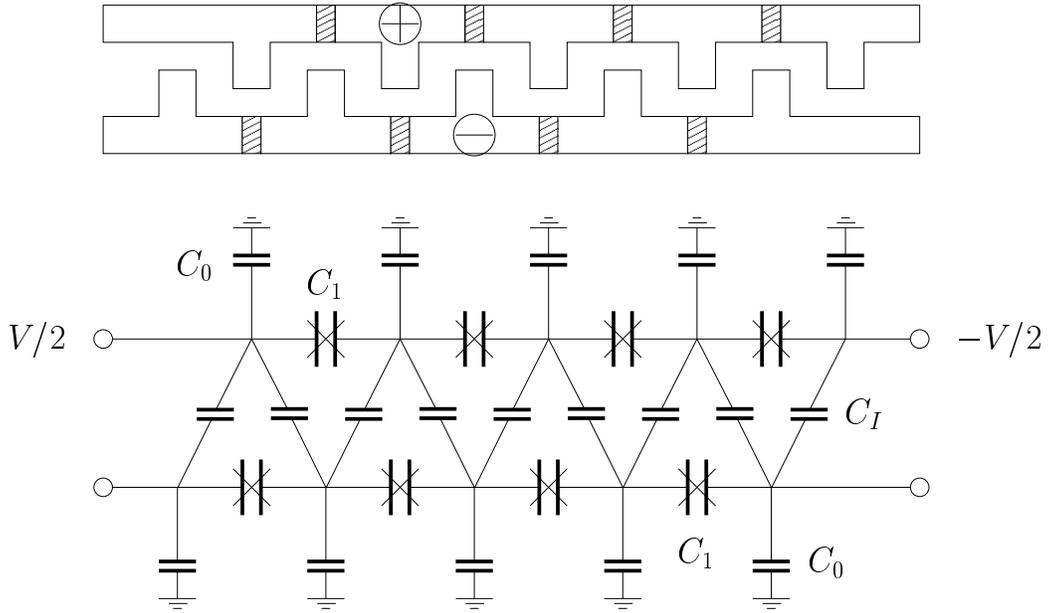,clip=,width=0.9\columnwidth}
\end{center}
\caption{Schematic picture and equivalent circuit of the system with
slanted inter-chain coupling. 
After Delsing, Haviland, and Davidsson, Czech. J. Phys. {\bf 46},
2359 (1996).}
\label{fig:2}
\end{figure}

Each of the two chains ($\ell=1,2$) of Josephson junctions considered
here is characterized by the Josephson coupling energy $E_J$ and the
charging energies $E_0{\equiv}e^2/2C_0$ and $E_1{\equiv}e^2/2C_1$
associated with the self-capacitance $C_0$ and the junction
capacitance $C_1$, respectively (see Fig.~\ref{fig:1} and
Fig.~\ref{fig:2}).
The two chains are coupled with each other by the capacitance $C_I$,
with which the electrostatic energy $E_I{\equiv}e^2/2C_I$ is
associated.
Two different ways of coupling are considered: 
Each island in one array is coupled one-to-one to one island (straight
coupling; Fig.~\ref{fig:1}) or aslant to two islands (slanted
coupling; Fig.~\ref{fig:2}) in the other array. 
There is no Cooper-pair tunneling between the chains.
The intra-chain capacitances are assumed to be so small
($E_0,E_1\gg{}E_J$) that, without the coupling, both chains
would each be in the insulating phase~\cite{Faziox91}.
It is also assumed that the coupling capacitance is sufficiently large
compared with the intra-array capacitances: $C_I\gg{}C_0,C_1$.
In that case, the electrostatic energy of the particle-hole pair
($\sim{}E_I$) is much smaller than that of an unpaired charge
($\sim{}E_0,E_1$).
For the most part, this work is devoted to the case of identical
chains, but non-identical chains will also be briefly discussed.

In the specified region of parameter space, the time scale of the
relevant dynamics is determined by the coupling capacitance ($C_I$)
and the corresponding Josephson plasma frequency
(${\sim}\sqrt{8E_IE_J}/\hbar$).
To be explicit, for straight coupling, it is convenient to rescale the
capacitances by $2C_I$ (i.e., $C_0/2C_I{\to}C_0,C_1/2C_I{\to}C_1$),
the energies by $\hbar\omega_p\equiv\sqrt{4E_IE_J}$, and the times by
$1/\omega_p$.
For slanted coupling, on the other hand, a more convenient choice is a
capacitance scale of $4C_I$ and a frequency scale of
$\hbar\omega_p\equiv\sqrt{2E_IE_J}$.
Although it is not essential, throughout this paper, I adopt these
normalizations of the physical quantities for simplicity of notation.

Then, the system with straight coupling can be well described by the
Hamiltonian (in units of $\hbar\omega_p$)
\begin{eqnarray}
H
& = & \frac{1}{2\sqrt{2}K}\sum_{\ell,\ell'}\sum_{x,x'}
  n(\ell;x)\CC^{-1}(\ell,\ell';x,x')
  n(\ell;x')
  \nonumber \\
& & \mbox{}
  - \sqrt{2}K\sum_\ell\sum_{<xx'>}
    \cos\left[\phi(\ell;x)-\phi(\ell;x')\right]
  ,
  \label{CCJJC:Ha}
\end{eqnarray}
where the coupling constant has been defined by
$K{\equiv}\sqrt{E_J/8E_I}$.
The number $n(\ell;x)$ of excess Cooper pairs and the phase
$\phi(\ell;x)$ of
the superconducting order parameter on the grain at $x$ in
the chain $\ell$ are quantum-mechanically conjugate variables:
$
[n(\ell;x),\phi(\ell';x')]
{=}i\delta_{xx'}\delta_{\ell\ell'}
$.
The Fourier transform of the capacitance matrix in Eq.~(\ref{CCJJC:Ha})
takes the following form (in units of $2C_I$):
\begin{equation}
\widetilde\CC(q)
= \tilC(q)\left( \begin{array}{cc}
    1 & 0 \\ 0 & 1
  \end{array} \right)
  + \frac{1}{2}\left( \begin{array}{rr}
    1 & -1 \\ -1 & 1
  \end{array} \right) ,
  \label{CC:1}
\end{equation}
where $\widetilde{C}(q)\equiv{}C_0{+}C_1\Delta(q)$, with
$\Delta(q)\equiv{}2(1{-}\cos{q})$, is the Fourier transform of the
submatrix $C(x,x')$ within either of the chains.
For slanted coupling, on the other hand, the appropriate
Hamiltonian reads as
\begin{eqnarray}
H
& = & \frac{1}{4K}\sum_{\ell,\ell'}\sum_{x,x'}
  n(\ell;x)\CC^{-1}(\ell,\ell';x,x')
  n(\ell;x')
  \nonumber \\
& & \mbox{}
  - 2K\sum_\ell\sum_{<xx'>}
    \cos\left[\phi(\ell;x)-\phi(\ell;x')\right]
  .
  \label{CCJJC:Hb}
\end{eqnarray}
In this case, the capacitance matrix (in units of $4C_I$) is
given by
\begin{eqnarray}
\widetilde\CC(q)
& = & \tilC(q)\left( \begin{array}{cc}
    1 & 0 \\ 0 & 1
  \end{array} \right)
  + \frac{1}{4}\left( \begin{array}{rr}
    1 & -1 \\ -1 & 1
  \end{array} \right)
  \nonumber \\
& & \mbox{}
  + \frac{1}{4}\left( \begin{array}{cc}
    1 & -e^{+iq} \\ -e^{-iq} & 1
  \end{array} \right)
  .
  \label{CC:2}
\end{eqnarray}

\section{2D Classical Vortex Representations}\label{sec:2DVG}

In this section, I transform each of the models in
Eq.~(\ref{CCJJC:Ha}) and Eq.~(\ref{CCJJC:Hb}) into equivalent 2D
classical system of vortices.  The resulting system reveals clearly
the nature of the phase transition that will be discussed in
Section~\ref{sec:QPT}.

\subsection{Straight Coupling}

It is convenient to write the partition function of the system in
the imaginary-time path-integral representation as
\begin{equation}
Z
= \prod_{\ell,x,\tau}
  \sum_{n(\ell;x,\tau)}
  \int_0^{2\pi}d\phi(\ell;x,\tau)\;
  \exp\left[ -\varS \right]
  \label{CCJJC:Z}
\end{equation}
with the Euclidean action
\begin{eqnarray}
\varS
& = & \frac{1}{2\sqrt{2}K}\sum_{\ell,\ell'}\sum_{x,x',\tau}
  n(\ell;x,\tau)\CC^{-1}(\ell,\ell';x,x')
  n(\ell;x',\tau)
  \nonumber \\
& & \mbox{}
  - \sqrt{2}K\sum_{\ell}\sum_{x,\tau}
    \cos\nabla_x\phi(\ell;x,\tau)
  \nonumber \\
& & \mbox{}
  + i\sum_{\ell}\sum_{x} n(\ell;x,\tau)
    \nabla_\tau\phi(\ell;x,\tau)
  ,
  \label{CCJJC:S}
\end{eqnarray}
where $\nabla_x$ and $\nabla_\tau$ denote the difference operators
with respect to $x$ and $\tau$, respectively, and the
(imaginary-)time slice $\delta\tau$ has been chosen to be unity (in
units of $1/\omega_p$)~\cite{end_note:2}.
The highly symmetric form of Eq.~(\ref{CCJJC:S}) with respect to space
and (imaginary) time makes it useful to introduce the space-time
2-vector notation $\bfr\equiv(x,\tau)$ and analogous notations for
all other vector variables.
We then apply the Villain approximation~\cite{Josexx77} to rewrite the
cosine term as summation over an integer field $\{m_x(\ell;\bfr)\}$.
Further, with the aid of the Poisson resummation
formula~\cite{Josexx77} and Gaussian integration, we also rewrite the
charging energy term as a summation over another integer field
$\{m_\tau(\ell;\bfr)\}$ to obtain the partition function
\begin{equation}
Z
\sim \prod_{\ell;\bfr}\sum_{\bfm(\ell;\bfr)}
  \int_{-\infty}^\infty d\phi(\ell;\bfr)\;
  \exp\left\{ -\varS \right\}
  \label{tmp:1a}
\end{equation}
with
\begin{eqnarray}
\varS
& = & \frac{K}{\sqrt{2}}\sum_{\ell,\ell';\bfr,\bfr'}
    \CC(\ell,\ell';x,x')
    \delta_{\tau\tau'}\;
    \left[
      \nabla_\tau\phi(\ell;\bfr)
      - 2\pi m_\tau(\ell;\bfr)
    \right]
    \left[
      \nabla_\tau\phi(\ell';\bfr')
      - 2\pi m_\tau(\ell';\bfr')
    \right]
  \nonumber \\
& & \mbox{}
  + \frac{K}{\sqrt{2}}\sum_{\ell;\bfr}
    \left[
      \nabla_x\phi(\ell;\bfr)-2\pi m_x(\ell;\bfr)
    \right]^2
  .
  \label{tmp:1b}
\end{eqnarray}
The variables
$\phi(\ell;\bfr)$ and $\bfm(\ell;\bfr)$
can be conveniently replaced by 
$\phi^\pm(\bfr)\equiv\phi(1;\bfr)\pm\phi(2;\bfr)$ and
$\bfm^\pm(\bfr)\equiv\bfm(1;\bfr)\pm\bfm(2;\bfr)$,
respectively.
In this way, one decomposes the Euclidean action in Eq.~(\ref{tmp:1b})
into the sum $\varS=\varS^++\varS^-$ with $\varS^\pm$ defined by
\begin{eqnarray}
\varS^\pm
& = &
  + \frac{K}{2\sqrt{2}}\sum_{\bfr,\bfr'}
    C^\pm(x,x')\delta(\tau,\tau')\;
    \left[
      \nabla_\tau\phi^\pm(\bfr)
      - 2\pi m_\tau^\pm(\bfr)
    \right]
    \left[
      \nabla_\tau\phi^\pm(\bfr')
      - 2\pi m_\tau^\pm(\bfr')
    \right]
  \nonumber \\
& & \mbox{}
  + \frac{K}{2\sqrt{2}}\sum_{\bfr}
    \left[
      \nabla_x\phi^\pm(\bfr)
      - 2\pi m_x^\pm(\bfr)
    \right]^2
  .
  \label{S:pm}
\end{eqnarray}
Here, the new capacitance matrices $C^\pm(x,x')$ have been defined by
$\tilC^+(q)=\tilC(q)$ and $\tilC^-(q)=1+\tilC(q)$.
Now, one follows the standard procedures~\cite{Korshu90,Faziox91} to
integrate out $\{\phi^\pm(\bfr)\}$.
Apart from the irrelevant spin wave part, one can finally obtain
the 2D system of classical vortices, which is also decomposed into
two subsystems $H_V=H_V^++H_V^-$ with
\begin{equation}
H_V^\pm
= \sqrt{2}\pi^2 K\sum_{\bfr,\bfr'}
    v^\pm(\bfr)\;U^\pm(\bfr-\bfr')\;
    v^\pm(\bfr')
    .
    \label{2DVG:H1}
\end{equation}
where the interactions between vortices
are defined via their Fourier transforms
\begin{eqnarray}
\tilU^\pm(\bfq)
= \frac{\tilC^\pm(q)}
  {\Delta(q)+\tilC^\pm(q)\Delta(\omega)}
  .
  \label{2DVG:U1}
\end{eqnarray}

\begin{figure}
\begin{center}
\epsfig{file=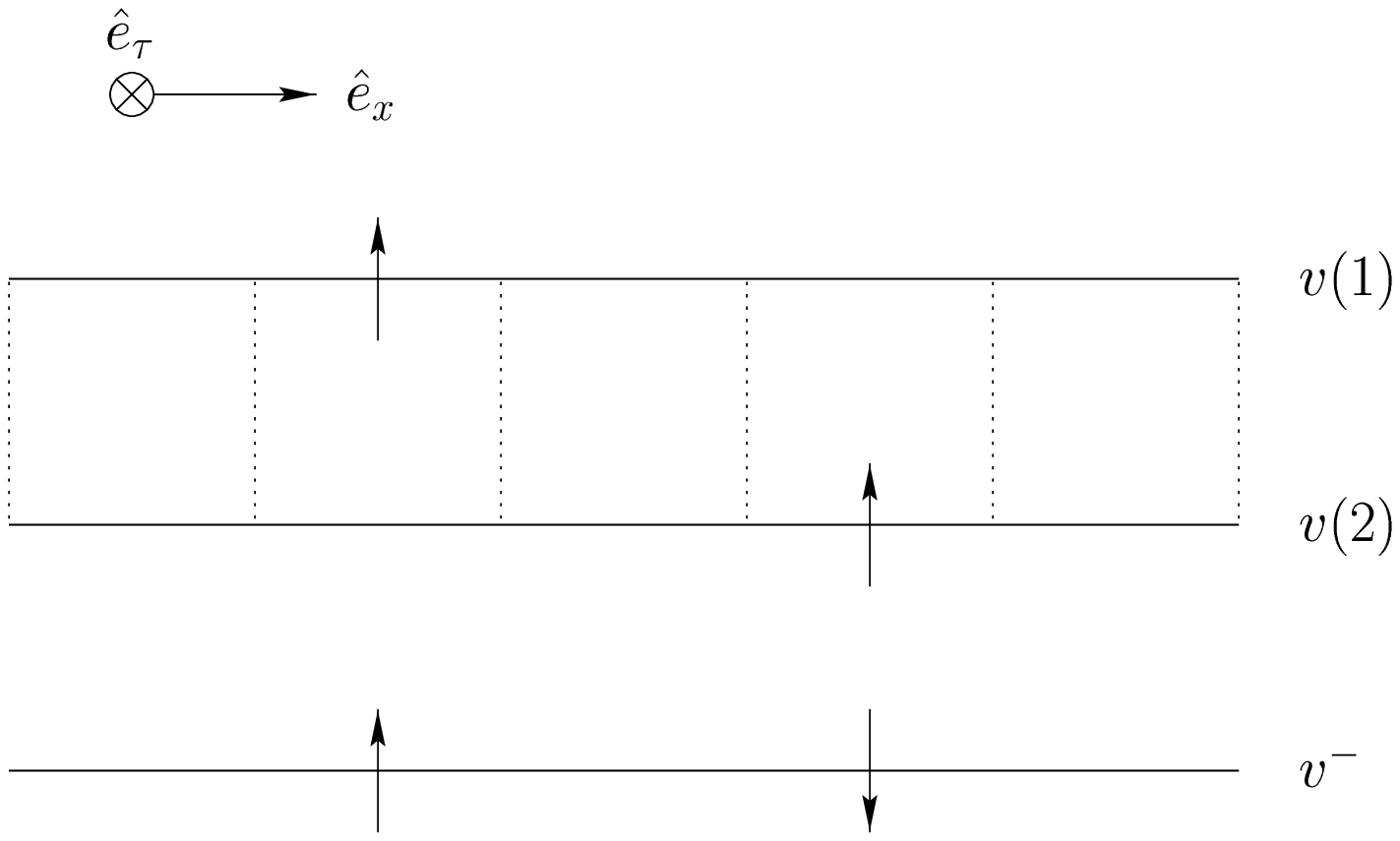,clip=,width=0.9\columnwidth}
\end{center}
\caption{Displacement vortex $v^-$. The configuration of one vortex
$v(1;\bfr){=}{+}1$ on space-time layer $1$ and another
$v(2;\bfr'){=}{+}1$ on layer $2$ ($\bfr \neq\bfr'$) corresponds to a
pair of displacement vortices, a vortex $v^-(\bfr){=}{+}1$ and an
antivortex $v^-(\bfr'){=}{-}1$. }
\label{fig:3}
\end{figure}

It is instructive to notice that the vortices $v^\pm$ represent
some sorts of correlations between the two chains in the system.
In Fig.~\ref{fig:3} is represented a configuration of two vortices,
$v(1;\bfr)=+1$ and $v(2;\bfr')=+1$ ($\bfr\neq\bfr'$).
This configuration gives a pair of a vortex $v^-(\bfr)=+1$ and an
antivortex $v^-(\bfr')=-1$, which tend to be bound to each other.
At the same time, for $v^+$, it gives two vortices
($v^+(\bfr)=v^+(\bfr')=+1$) which are inclined to repel each other.
For the configuration with a vortex on one space-time layer and
an antivortex on the other, we have opposite tendencies for $v^+$ and
$v^-$.
%
%
As I will show in Section~\ref{sec:QPT}, it is the vortices $v^-$ that
play major roles in the quantum phase transition of the system.
Furthermore, the vortices $v^-$ will be shown to be manifestations of
the particle-hole pairs.

\begin{figure}
\centerline{\epsfig{file=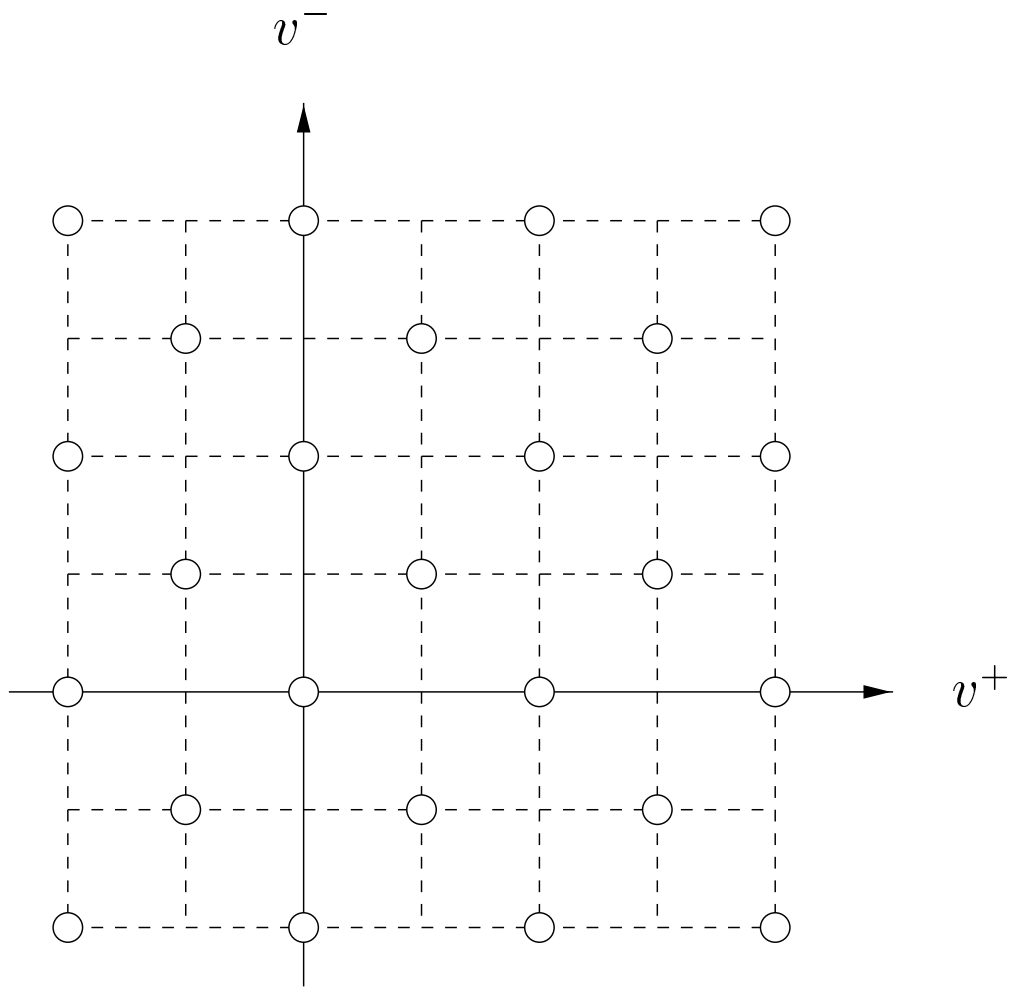,clip=,width=0.9\columnwidth}}
\caption{Topological coupling of $v^+$ and $v^-$.  At each
space-time position $\vec\bfr$, $(v^+,v^-)$ can take only half
of the elements in ${\bf{}Z{\times}Z}$ as depicted with the open circles
in the figure.}
\label{fig:4}
\end{figure}

In Eq.~(\ref{2DVG:H1}), $U^-(0)$ always diverges and gives rise to the
{\em vortex number equality condition}
$\sum_\bfr{}v(1;\bfr)=\sum_\bfr{}v(2;\bfr)$, or equivalently, the
vorticity neutrality condition $\sum_\bfr{}v^-(\bfr)=0$ for $v^-$.
A similar neutrality condition, $\sum_\bfr{}v^+=0$, should be satisfied
unless $C_0=0$ (see Ref.~\onlinecite{mschoi98a}).
More importantly, it should be noticed that the two fields
$v^+$ and $v^-$ actually are not independent of each other since
$m_\mu(1;\bfr)$ and $m_\mu(2;\bfr)$ in Eq.~(\ref{S:pm}), and
hence $v(1;\bfr)=[v^+(\bfr)+v^-(\bfr)]/2$
and $v(2;\bfr)=[v^+(\bfr)-v^-(\bfr)]/2$, can
take only integer values.
As depicted with open circles in Fig.~\ref{fig:4}, $(v^+,v^-)$ at each
$\bfr$ can take only half of the elements in the product set of
integers ${\bf{}Z{\times}Z}$; $v^+$ and $v^-$ are {\em topologically
coupled} with each other.
However, this topological coupling is irrelevant and can be safely
neglected, which will be discussed in more detail in
Section~\ref{sec:QPT}.

\subsection{Slanted Coupling}

\begin{figure}
\centerline{\epsfig{file=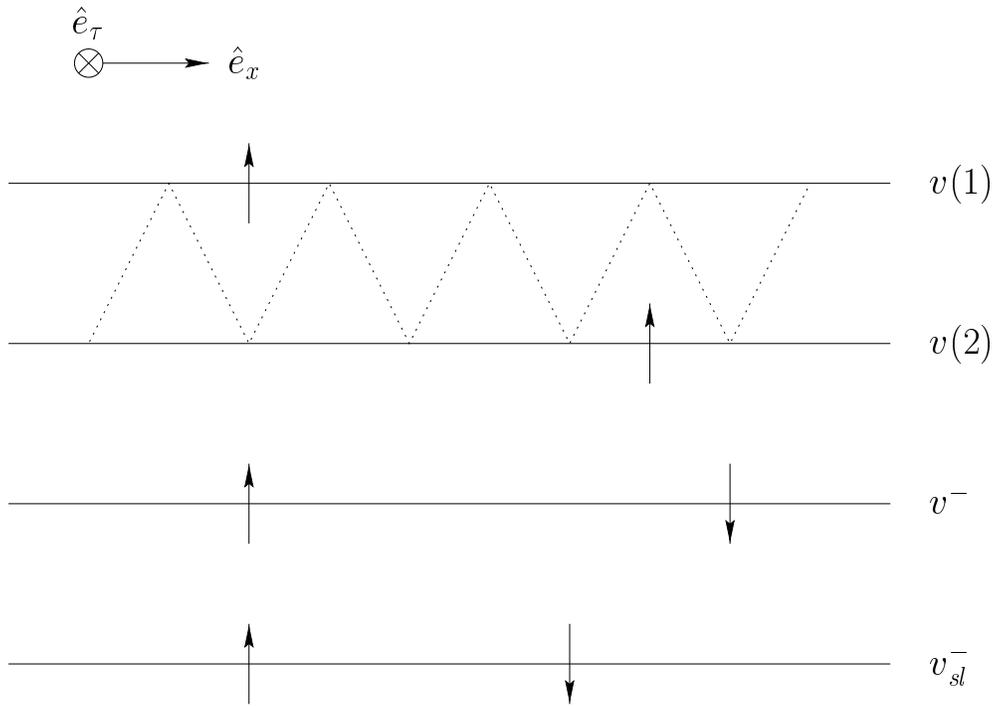,clip=,width=0.9\columnwidth}}
\caption{Displacement vortex for slanted coupling.  At long
(space-time) lengths, the vortex-antivortex pair of $v^-_{\it{}sl}$
gives the same energy as that of $v^-$. }
\label{fig:5}
\end{figure}

In the same manner as that leading to Eq.~(\ref{tmp:1b}), from the
path-integral representation of the partition function
Eq.~(\ref{CCJJC:Z}), one can obtain the partition function for slanted
coupling:
\begin{equation}
Z
\sim \prod_{\ell;\bfr}\sum_{\bfm(\ell;\bfr)}
  \int_{-\infty}^\infty d\phi(\ell;\bfr)\;
  \exp\left\{ -\varS \right\}
  \label{tmp:2a}
\end{equation}
with
\begin{eqnarray}
\varS
& = & K\sum_{\ell,\ell';\bfr,\bfr'}
    \CC(\ell,\ell';x,x')
    \delta_{\tau\tau'}\;
    \left[
      \nabla_\tau\phi(\ell;\bfr)
      - 2\pi m_\tau(\ell;\bfr)
    \right]
    \left[
      \nabla_\tau\phi(\ell';\bfr')
      - 2\pi m_\tau(\ell';\bfr')
    \right]
  \nonumber \\
& & \mbox{}
  + K\sum_{\ell;\bfr}
    \left[
      \nabla_x\phi(\ell;\bfr)-2\pi m_x(\ell;\bfr)
    \right]^2
  .
  \label{tmp:2b}
\end{eqnarray}
Unlike the previous case, owing to the last term in the capacitance
matrix $\CC$ in Eq.~(\ref{CC:1}), the replacement of the variables
$\phi(\ell;\bfr)$ and $\bfm(\ell;\bfr)$ by $\phi^\pm(\bfr)$ and
$\bfm^\pm(\bfr)$, respectively, provides no help at this stage.
Instead, we directly integrate out $\phi(\ell;\bfr)$ to get the
Hamiltonian for 2D classical vortices:
\begin{eqnarray}
H_V
& = & 4\pi^2K\sum_{\ell;\bfr,\bfr'}
    v(\ell;\bfr)U_0(\bfr,\bfr')v(\ell;\bfr')
  \nonumber \\
& + & 4\pi^2K\sum_{\bfr,\bfr'}
    v^-(\bfr)
    U_I(\bfr,\bfr')
    v^-(\bfr')
  \nonumber \\
& + & 4\pi^2K\sum_{\bfr,\bfr'}
    v^-_{\it sl}(\bfr)
    U_I(\bfr,\bfr')
    v^-_{\it sl}(\bfr')
    ,
  \label{2DVG:H2a}
\end{eqnarray}
where the vortex interactions are given via the Fourier transforms
\begin{eqnarray}
\tilU_0(\bfq)
& = & \frac{1}{T(\bfq)}\left[
    \tilC(q)\left[
      \Delta(q) + \left\{1+\tilC(q)\right\}\Delta(\omega)
    \right]
    + \frac{1}{16}\Delta(q)\Delta(\omega)
  \right]
  ,
  \\
\tilU_I(\bfq)
& = & \frac{1}{T(\bfq)}\frac{\Delta(q)}{4}
\end{eqnarray}
with
\begin{equation}
T(\bfq)
= \left[
      \Delta(q) + \tilC(q)\Delta(\omega)
    \right] \left[
      \Delta(q) + \left\{1+\tilC(q)\right\}\Delta(\omega)
    \right]
  + \frac{1}{16}\Delta(q)\Delta^2(\omega)
  .
  \label{tmp:3}
\end{equation}
In Eq.~(\ref{2DVG:H2a}), the vortices $v^-$ and $v^-_{\it{}sl}$ are
defined by the direct differences
$v^-\equiv\left[v(1;\bfr)-v(2;\bfr)\right]$ and by the slanted
differences
$v^-_{\it{}sl}\equiv\left[v(1;\bfr)-v(2;\bfr{+}\hate_x)\right]$
(see Fig.~\ref{fig:5}).
The appearance of the additional term in $v^-_{\it{}sl}$ in
Eq.~(\ref{2DVG:H2a}) is no surprise since with a vortex on one
space-time layer the correspondent vortex has two alternative
nearest-neighbor sites to sit on on the other space-time layer.
However, at the long times and lengths that we are
interested in, any configuration of $v^-_{\it{}sl}$ gives the
same energy as the corresponding configuration of $v^-$,
which can be seen in Fig.~\ref{fig:5}.
Thus, $v^-_{\it{}sl}$ in the last additional term in
Eq.~(\ref{2DVG:H2a}) can be replaced simply by $v^-$, leading to
\begin{eqnarray}
H_V
& \simeq & 4\pi^2K\sum_{\ell;\bfr,\bfr'}
    v(\ell;\bfr)U_0(\bfr,\bfr')v(\ell;\bfr')
  \nonumber \\
& + & 8\pi^2K\sum_{\bfr,\bfr'}
    v^-(\bfr)
    U_I(\bfr,\bfr')
    v^-(\bfr')
  .
  \label{2DVG:H2c}
\end{eqnarray}

This point can also be seen in a more rigorous way by rewriting the
Hamiltonian in Eq.~(\ref{2DVG:H2a}) in terms of $v^\pm(\bfr)$; i.e.,
$H_V=H_V^++H_V^-+\delta{H}_V$, where $H_V^\pm$ has the anticipated form
\begin{equation}
H_V^\pm
= 2\pi^2K\sum_{\bfr,\bfr'}
  v^\pm(\bfr)\: U^\pm(\bfr,\bfr')\: v^\pm(\bfr')
  \label{2DVG:H2b}
\end{equation}
with the vortex interactions
\begin{eqnarray}
\tilU^+(\bfq)
& = & \frac{1}{T(\bfq)}\left[
    \tilC(q)\left[
      \Delta(q) + \left\{1+\tilC(q)\right\}\Delta(\omega)
    \right]
    + \frac{1}{16}\Delta(q)\Delta(\omega)
  \right]
  \label{tmp:4a}
  \\
\tilU^-(\bfq)
& = & \frac{1}{T(\bfq)}\left[
    \left\{1+\tilC(q)\right\}\left\{
      \Delta(q) + \tilC(q)\Delta(\omega)
    \right\}
    + \frac{1}{16}\Delta(q)\Delta(\omega)
  \right]
  .
  \label{tmp:4b}
\end{eqnarray}
The term $\delta{H}_V$,
defined by
\begin{equation}
\delta{H}_V
= \frac{\pi^2}{4}K\sum_{\alpha,\beta=\pm}\sum_{\bfr,\bfr'}
  v^\alpha(\bfr)\:\delta{U}^{\alpha\beta}\: v^\beta(\bfr')
\end{equation}
with 
\begin{equation}
\delta\tilde{U}^{\alpha\beta}(\bfq)
= \frac{\Delta(q)}{T(\bfq)}\left( \begin{array}{cc}
    +\Delta(q) & +2i\sin{q} \\ -2i\sin{q} & -\Delta(q)
  \end{array}\right)
  ,
\end{equation}
describes the interaction between $v^+$ and
$v^-$.  Since the numerator in the interaction
$\delta\tilU^{\alpha\beta}(\bfq)$ is the third order in $\bfq$,
at long times and lengths ($\bfq\to0$), $\delta{H}_V$ can be ignored
compared with $H_V^\pm$.
Moreover, in the low frequency and
momentum limit (neglecting the terms of order ${\cal{}O}(\bfq^4)$
or higher, the vortex interactions in
Eqs.~(\ref{tmp:4a}) and (\ref{tmp:4b}) are simply reduced to those in
Eq.~(\ref{2DVG:U1}) for straight coupling. 
For this reason, it is concluded that, at long times and lengths,
the 2D vortex representation for slanted coupling is equal to
that for straight coupling except for the different coupling
constant: $K/\sqrt{2}$ for straight coupling and $K$ for
slanted coupling.

\section{Linear Response}\label{sec:sigma}

The mathematical mapping in the previous section enables us to examine
the existence and the universal class of the phase transition, and yet
we need to identify and characterize the phases on both sides of
the transition (see Section \ref{sec:QPT}).
A common method is to measure the response of the
system to an external perturbation.
In this section, I consider the conductivity, specifically, the linear
response $\sigma_{\ell_1\ell_2}(\omega)$ of the current in chain
$\ell_1$ to the voltage applied across chain $\ell_2$ (see
Fig.~\ref{fig:1} and Fig.~\ref{fig:2}) and rewrite it in the vortex
representation in accordance with the previous section. 

The response function $\sigma_{\ell_1\ell_2}(\omega)$ can be obtained
via the analytic continuation
\begin{equation}
\sigma_{\ell_1\ell_2}(\omega)
= \frac{1}{i\omega}\lim_{q\to0}
  \widetilde\varG_{\ell_1\ell_2}(q,i\omega'\to\omega+i0^+),
  \label{sigma:def}
\end{equation}
where $\widetilde\varG_{\ell_1\ell_2}$ is the Fourier transform of the
imaginary-time Green's function
\begin{equation}
\varG_{\ell_1\ell_2}(x,\tau)
= \avgl T_\tau [I(\ell_1;x,\tau)I(\ell_2;0,0)] \avgr
  \label{varG:def}
\end{equation}
with the time-ordered product $T_\tau$
and the current operators
$I(\ell;x)\equiv\sin\nabla_x\phi(\ell;x)$.
Owing to the symmetry between the two chains, the conductivities
$\sigma_{11}$ and $\sigma_{21}$ can be written as
\begin{eqnarray}
\sigma_{11}(\omega)
& = & \frac{1}{4}\left[\sigma_+(\omega)+\sigma_-(\omega)\right]
  , \\
\sigma_{21}(\omega)
& = & \frac{1}{4}\left[\sigma_+(\omega)-\sigma_-(\omega)\right]
\end{eqnarray}
in terms of $\sigma_\pm$ defined in a manner analogous to
Eqs.~(\ref{sigma:def}) and (\ref{varG:def}) with
$I^\pm(x)\equiv{}I(1;x)\pm{}I(2;x)$.
In the same manner as in Section \ref{sec:model}, one can get the
vortex representation of the corresponding Green's functions
$\varG_\pm$ (see Appendix~\ref{appx:A}):
\begin{equation}
\varG_\pm(\bfr_1,\bfr_2)
= \nabla_{\tau_1}\nabla_{\tau_2}\left\{
    -\frac{\gamma}{K}U^\pm(\bfr_1,\bfr_2)
    + 4\pi^2\sum_{\bfr_1',\bfr_2'}
      U^\pm(\bfr_1,\bfr_1')U^\pm(\bfr_2,\bfr_2')
      \avgl v^\pm(\bfr_1')v^\pm(\bfr_2') \avgr_V
  \right\}
  ,
  \label{varGpm}
\end{equation}
where $\gamma=\sqrt{2}$ for straight coupling or $\gamma=1$ for
slanted coupling, and the average $\avgl\cdots\avgr_V$ is with
respect to the total vortex Hamiltonian $H_V=H_V^++H_V^-$.
In Eq.~(\ref{varGpm}), the same Hamiltonians $H_V^\pm$ as in
Eq.~(\ref{2DVG:H1}) and the same interactions $U^\pm$ as in
Eq.~(\ref{2DVG:U1}) are used both for straight and slanted
coupling.
This should be valid at the long times and lengths we are interested
in, as discussed in the previous section.
One of the benefits of the representation in Eq.~\ref{varGpm} is that
the vortex contribution of the second term in Eq.~\ref{varGpm} can be
estimated in the standard renormalization group (RG) approach for 2D
classical vortices~\cite{Bradle84,Koster74}.

\section{Quantum Phase Transitions}\label{sec:QPT}

Now, I turn to the main subjects of this work, the quantum-phase
transition and the current-drag effect in the system based on the 2D
classical-vortex representations of the partition functions
in Eq.~(\ref{2DVG:H1}) and in Eq.~(\ref{2DVG:H2b}) and the response
functions in Eq.~\ref{sigma:def}.
In Section \ref{sec:model}, it was established that, aside from the
differences in the coupling constant, the two coupling schemes are
equivalentat at long times and lengths.
First, the discussion will focuse on straight coupling,
the results of which can be applied in full to slanted coupling with
the coupling constant properly replaced.
Some important differences between the two coupling schemes will be
discussed at the end of the section.

It is not difficult to understand separately the physics described by
each of the Hamiltonians $H_V^\pm$ in Eq.~(\ref{2DVG:H1}).
Unless $C_0=0$, the length-dependent anisotropy due to
$\tilC^+(q)=\tilC(q)\ll1$ in $\tilU^+(\bfq)$ in Eq.~(\ref{2DVG:U1})
fades out on (space-time) length scales larger than $\sqrt{C_1/C_0}$,
and thereby $\hatU^+(\bfr)\equiv{}2\pi[U^+(0)-U^+(\bfr)]$ is simply
reduced to the isotropic logarithmic interaction
\begin{equation}
\hatU^+(\bfr)
\simeq \sqrt{C_0}\log{r} ,\quad (r\gg 1)
  .
\end{equation}
This results in the usual vortex Hamiltonian
\begin{equation}
H_V^+
\simeq -\pi K_{\it eff}^+\sum_{\bfr,\bfr'}
  v^+(\bfr)\log|\bfr-\bfr'|v^+(\bfr)
  \label{tmp:5a}
\end{equation}
with the effective coupling constant
$K_{\it{}eff}^+\equiv\sqrt{E_J/16E_0}$.
Because we assumed at the beginning of the paper that
$E_0,E_1\gg{}E_J$, $K_{\it{}eff}$ is substantially smaller than the
BKT transition point $K_{\it{}BKT}\simeq{}2/\pi$; the vortices $v^+$
always form a neutral plasma of free vortices regardless of $K$ (i.e.,
regardless of $C_I$).
In the case of $C_0=0$, $U^+(\bfr)$ is only short ranged:
$U^+(\bfr)\sim\sqrt{C_1}\exp(-r/\sqrt{C_1})$.  In this case, the
vortices $v^+$ even form a non-neutral plasma of free vortices.
In any case, the system of vortices $v^+$ always forms a plasma of
free vortices, regardless of the value of $K$.
On the other hand, neglecting the short-distance anisotropy,
$\hatU^-(\bfr)$ is isotropic in space-time and logarithmic in
distance: $\tilU^-(\bfq)\simeq{}1/[\Delta(q)+\Delta(\omega)]$ and
$\hatU(\bfr)\simeq\log{r}$ leading to the 2D Coulomb gas Hamiltonian
for $v^-$
\begin{equation}
H_V^-
\simeq -\sqrt{2}\pi^2K\sum_{\bfr,\bfr'}
  v^-(\bfr)\log|\bfr-\bfr'|v^-(\bfr')
  .
  \label{tmp:5b}
\end{equation}
It follows that the system of vortices $v^-$ exhibits a BKT-type phase
transition at $K=K_c^-$ such that
$K_c^-/\sqrt{2}=K_{\it{}BKT}\simeq{}2/\pi$.

At this point, one might be tempted to conclude that, as $K$ is
decreased, the total system $H_V$ goes through a BKT-type
transition at $\sqrt{2}K_{\it{}BKT}$ which is entirely driven by the
vortices $v^-$, with $v^+$ playing no role.
This scenario, however, should be carefully tested against the
topological coupling between $v^+$ and $v^-$ discussed in the previous
section.

\begin{figure}
\centerline{\epsfig{file=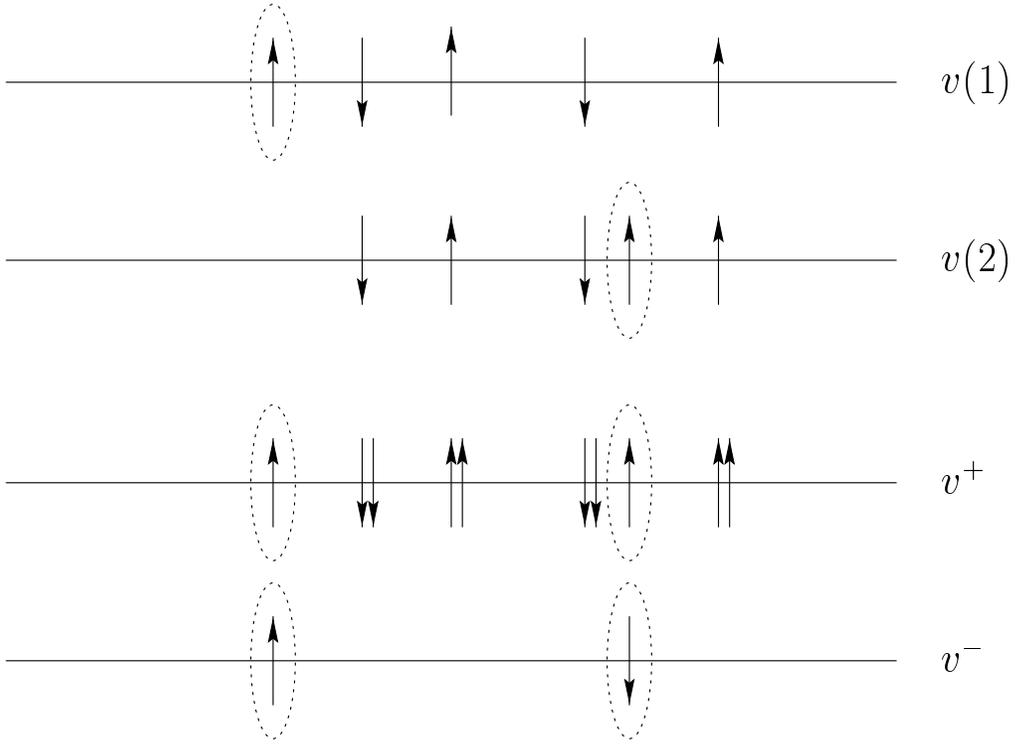,clip=,width=0.9\columnwidth}}
\caption{Due to the topological coupling, the vortex-antivortex pair
of the displacement vortices $v^-$ (indicated by dotted ellipses)
induces two vortices of $v^+$.  The interaction between these two
vortices is, however, completely screened out by the free vortices of
$v^+$ with $\pm2$ (indicated by the douple arrows).}
\label{fig:6}
\end{figure}

For this goal, it is convenient to consider the subsystem
$\{v^*_\mu\}$ of $\{v^+_\mu\}$ satisfying $v^-(\bfr)=0$ for all $\bfr$.
In this subsystem, $v^*(\bfr)$ can take only even numbers, and
hence the Hamiltonian can be written as
\begin{equation}
H_V^*
\simeq -\pi(4K_{\it{}eff}^+)\sum_{\bfr,\bfr'}
  [v^*(\bfr)/2]\log|\bfr-\bfr'|[v^*(\bfr')/2]
\end{equation}
unless $C_0=0$.  Under the assumption $E_0,E_1\gg{}E_J$
($4K_{\it{}eff}^+\ll{}K_{\it{}BKT}$),  there are a significant number
of free vortices of $v^*$.  Obviously, it is also true in the case of
$C_0=0$.  These free vortices of $v^*$ substantially affect the
topological coupling and eventually make it irrelevant:
As illustrated in Fig.~\ref{fig:6}, the vortex-antivortex pair of
$v^-$ is always accompanied by two vortices of $v^+$ (enclosed in
dotted ellipses).
The interaction between these two vortices of $v^+$ is completely
screened out by the free vortices of $v^*$ (indicated by double arrows
in the figure).
The two vortices of $v^+$, therefore, cannot affect the interaction
energy of the vortex-antivortex pair of $v^-$; they can only slightly
change the fugacity of $v^-$.  
It is concluded that, as $K$ is increased, {\em the system exhibits a
BKT-type transition at
$K_c^{\it{}st}\simeq{}\sqrt{2}K_{\it{}BKT}\simeq{}2\sqrt{2}/\pi$,
which is exclusively attributed to the vortices $v^-$.} 

\begin{figure}
\centerline{\epsfig{file=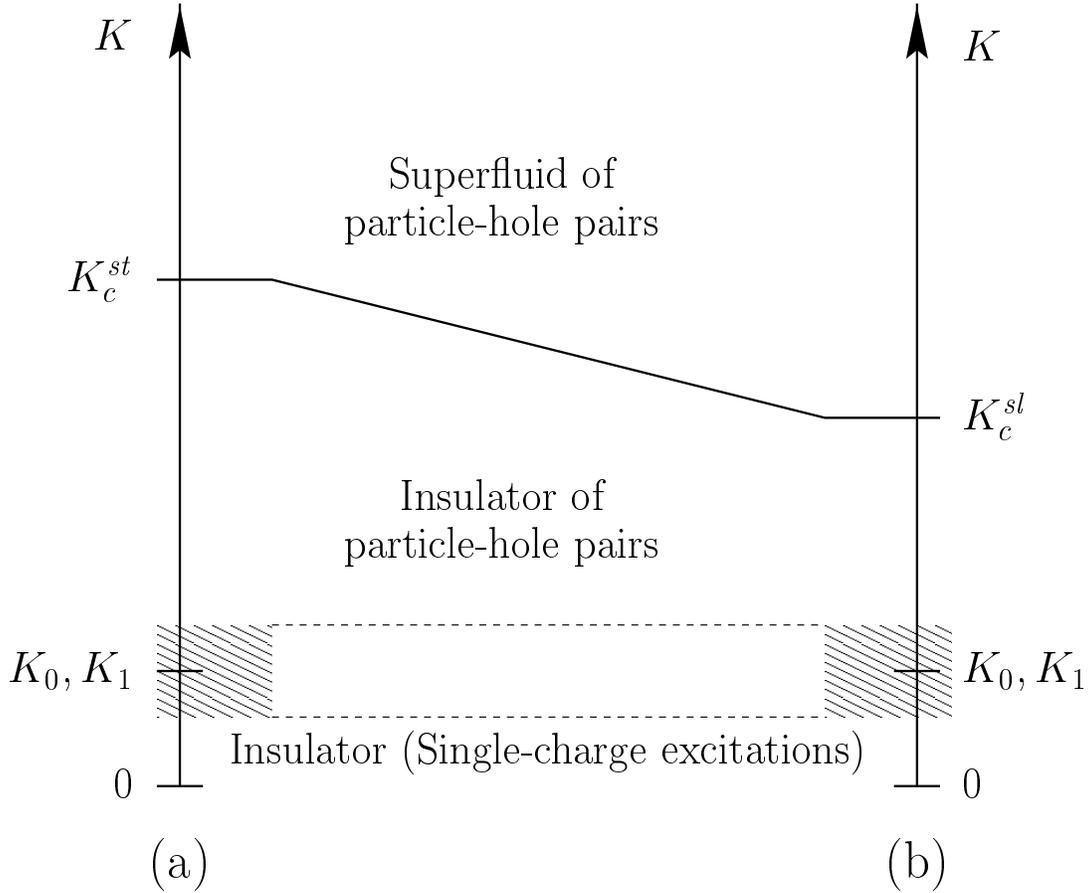,clip=,width=0.9\columnwidth}}
\caption{Phase diagrams of capacitively coupled Josephson-junction
chains (a) with straight coupling and (b) with slanted coupling.  The
intra-chain coupling constants $K_0$ and $K_1$ are defined by
$K_0\equiv\sqrt{E_J/8E_0}\propto\sqrt{C_0}$ and by
$K_1\equiv\sqrt{E_J/8E_1}\propto\sqrt{C_1}$, respectively.
In the text, only the region of $K\gg{}K_0,K_1$, where particle-hole
pairs are relevant, is considered.}
\label{fig:7}
\end{figure}

To examine the states of the system on both sides of $K_c^{\it{}st}$,
it is first noted that the vortices $v^\pm(\bfr)$ are topological
singularities in the 2D  space-time configurations of the phases
$\phi^\pm(\bfr)\equiv\phi(1;\bfr)\pm\phi(2;\bfr)$.
The plasma of free vortices of $v^+$ leads to complete disorder of
$\phi^+$.  Accordingly to the uncertainly relation between $\phi^+$
and the conjugate variable $n^+(x)\equiv[n(1;x)+n(2;x)]/2$, i.e.,
$\Delta\phi^+\Delta{}n^+\geq{}1$, this means that $n^+(x)$ for each
$x$ takes well-defined value, which should be zero due to the large
single-charge(Cooper pair) excitation energy of order of $E_0$ or
$E_1$: $n^+(x)=0$.
Consequently, the variable $n^-(x)\equiv[n(1;x)-n(2;x)]/2$ conjugate
to $\phi^-$ represents the number of pairs of excess and deficit
Cooper pairs (i.e., particle-hole pairs).  Further, it is also noted
that the effective model in Eq.~(\ref{2DVG:H1}) or Eq.~(\ref{tmp:5b})
is a vortex representation of the quantum phase
model~\cite{Bradle84,mschoi98a}
\begin{equation}
H_{\it QPM}
= \frac{1}{\sqrt{2}K}\sum_x\left[n^-(x)\right]^2
  + \frac{K}{\sqrt{2}}\sum_x\cos\nabla_x\phi^-(x) 
  .
\end{equation}
Therefore, the BKT-type phase transition at $K_c^{\it{}st}$ driven by
$v^-$ is nothing but an insulator-to-superfluid transition of the
particle-hole pairs:  Although particle-hole pairs are always the
lowest excitations, below $K_c^{\it{}st}$, they cannot move along the
system without external bias due to the Coulomb blockade associated
with the charging energy $E_I$.  For $K>K_c^{\it{}st}$, on the other
hand, the particle-hole pairs condensate to form a superfluid and can
move free along the system (see Fig.~\ref{fig:7}).
The formation of bound dipoles of vortices $v^-$ is an effective
manifestation of the condensation of the particle-hole pairs.

Such particle-hole transport can be confirmed by examining the current
responses the two chains given by Eq.~(\ref{sigma:def}).
Due to the free vortices of $v^+$, it follows
directly from Eq.~(\ref{varGpm}) that $\sigma_+(\omega)$ vanishes
($\omega\ll{}1$).  On the other hand, the tightly bound
vortex-antivortex pairs of $v^-$ above $K_c^{\it{}st}$ result in the
response function 
\begin{equation}
\sigma_-(\omega)
= \frac{2\pi}{\sqrt{2}K}\left(2-\frac{K}{K_R}\right)
  \delta(\omega)
  \quad (\omega \ll 1)
  ,
\end{equation}
and hence
\begin{equation}
\sigma_{11}(\omega)
= - \sigma_{21}(\omega)
= \frac{\pi}{2\sqrt{2}K}\left(2-\frac{K}{K_R}\right)
  \delta(\omega)
  ,
\end{equation}
where the renormalized coupling constant $K_R$ is defined by
\begin{equation}
\frac{\sqrt{2}}{K_R}
\equiv \frac{\sqrt{2}}{K} 
  - \frac{\pi^2}{2}\sum_\bfr
    |\bfr|^2 \avgl v^-(\bfr)v^-({\bf 0}) \avgr
  .
\end{equation}
Thus the system exhibits {\em superconductivity} and carries currents
along the two chains {\em equally large in magnitude but opposite in
direction}.  This perfect drag of supercurrents reveals that the
charges indeed transport in the form of particle-hole pairs, which are
bound by the electrostatic energy $E_I$ associated with $C_I$.
For $K<K_c^{\it{}st}$, on the other hand, the system
displays insulating particle-hole $I$-$V$ characteristics,
qualitatively the same as those in
Refs.~\cite{Averin91a,Matter97,Delsin96}.

The argument so far also holds for slanted coupling if only one
replaces $K/\sqrt{2}$ by $K$; the system with slanted coupling
exhibits a BKT-type transition at $K_c^{\it{}sl}\simeq{}K_{\it{}BKT}$,
and the superfluid state is characterized by the response functions
\begin{equation}
\sigma_{11}(\omega)
= - \sigma_{21}(\omega)
= \frac{\pi}{4K}\left(2-\frac{K}{K_R}\right)
  \delta(\omega)
  ,
\end{equation}
where
\begin{equation}
\frac{1}{K_R}
\equiv \frac{1}{K} 
  - \frac{\pi^2}{2}\sum_\bfr
    |\bfr|^2 \avgl v^-(\bfr)v^-({\bf 0}) \avgr
  .
\end{equation}
It is interesting, however, to notice that $K_c^{\it{}sl}$ is quite a
bit smaller than $K_c^{\it{}st}$ (see Fig.~\ref{fig:7}).  This
reflects the difference between the two coupling scheme in the
underlying transport mechanism: The correlated sequential tunneling of
particle-hole pairs, a first-order process, is more likely than
cotunneling, a second-order process.

\section{Conclusion}\label{sec:concl}

In conclusion, the properties associated with particle-hole pairs in
two capacitively coupled Josephson-junction chains, considering both
the straight and the slanted couplings have been investigated. 
In particular, the transport of particle-hole pairs was found
to drive the BKT-type insulator-to-superfluid transition with respect
to the coupling capacitance, regardless of the coupling scheme.  The
superfluid phase ($K>K_c$) is uniquely characterized by the absolute
drag of supercurrents along the two chains.

\section*{Acknowledgment}

I am grateful to M. Y. Choi, S.-I. Lee, and J.~V. Jos\'e for valuable
discussions and for sending me preprints.
This work was supported by the Minitry of Science and Technology of
Korea through the by the Creative Research Initiative Program.

\appendix
\section{}\label{appx:A}

In this appendix, the vortex-representation of the imaginary-time
Green's function in Eq.~(\ref{varGpm}) is derived.  For simplicity,
only the $\varG_-$ for straight coupling is derived here.  The
application of the same approach to $\varG_+$ should be
straightforward, though.  In addition, at long times and lengths, the
derivation should also hold for slanted coupling if one replace
$K/\sqrt{2}$ by $K$ (see Section~\ref{sec:2DVG}).

In the imaginary-time path-integral representation, the Green's
function
$\varG_-(\bfr_1,\bfr_2)\equiv\avgl{I_-(\bfr_1)I_-(\bfr_2)}\avgr$ can
be written as
\begin{eqnarray}
\varG_-(\bfr_1,\bfr_2)
& = &  \frac{1}{2K^2Z}\prod_{\ell;\bfr}
  \sum_{n(\ell;\bfr)}
  \int_0^{2\pi}d\phi(\ell;\bfr)\;
  \left\{
    \frac{\partial}{\partial\nabla_x\phi(1;\bfr_1)}
    - \frac{\partial}{\partial\nabla_x\phi(2;\bfr_1)}
  \right\}
  \nonumber \\
& & \mbox{}\times
  \left\{
    \frac{\partial}{\partial\nabla_x\phi(1;\bfr_2)}
    - \frac{\partial}{\partial\nabla_x\phi(2;\bfr_2)}
  \right\}
  \exp\left\{ -\varS[n,\phi] \right\}
  ,
\end{eqnarray}
where the Euclidean action $\varS$ is given by Eq.~(\ref{CCJJC:S}).
By changing the variables from $\phi(\ell;\bfr)$ and $\bfm(\ell;\bfr)$
to $\phi^\pm(\bfr)$ and $\bfm^\pm(\bfr)$, respectively, one obtains
\begin{equation}
\varG_-(\bfr_1,\bfr_2)
= \frac{2}{K^2Z}\prod_{\alpha=\pm}\prod_\bfr
  \sum_{\bfm^\alpha(\bfr)}
  \int d\phi^\alpha(\bfr)\;
  \frac{\partial}{\partial\nabla_x\phi^-(\bfr_1)}
  \frac{\partial}{\partial\nabla_x\phi^-(\bfr_2)}
  \exp\left\{ -\varS^+ -\varS^- \right\}
  .
\end{equation}
$\varS^\pm$ has been given in Eq.~(\ref{S:pm}).  The
$\phi^\pm$-integration can be performed easier by first introducing an
auxiliary field as follows:
\begin{equation}
\varG_-(\bfr_1,\bfr_2)
= -\frac{2}{K^2Z}\prod_{\alpha;\bfr}
  \sum_{\bfm^\alpha(\bfr)}
  \int d\phi^\alpha(\bfr)
  \int d^2\bfJ^\alpha(\bfr)\;
  J^-_x(\bfr_1)J^-_x(\bfr_2)
  \exp\left\{ -\varS^+-\varS^- \right\}
  ,
\end{equation}
where
\begin{eqnarray}
\varS^\pm
& = & \frac{1}{\sqrt{2}K}\sum_{\bfr,\bfr'} J_\tau^\pm(\bfr)\:
    \left[C^\pm(\bfr,\bfr')\right]^{-1}J_\tau^\pm(\bfr')
  + \frac{1}{\sqrt{2}K}\sum_{\bfr,\bfr'} \left[J_x^\pm(\bfr)\right]^2
  \nonumber \\
& & \mbox{}
  + i\sum_\bfr\bfJ(\bfr)\cdot\left[
      \nabla\phi^\pm(\bfr)-2\pi\bfm^\pm(\bfr)
    \right]
  .
\end{eqnarray}
Now, integrating out $\phi^\pm(\bfr)$, one finally gets the
vortex-representation of the Green's function $\varG_-$: 
\begin{equation}
\varG_-(\bfr_1,\bfr_2)
= \nabla_{\tau_1}\nabla_{\tau_2}\left\{
    -\frac{\sqrt{2}}{K}U^-(\bfr_1,\bfr_2)
    + 4\pi^2\sum_{\bfr_1',\bfr_2'}
      U^-(\bfr_1,\bfr_1')U^-(\bfr_2,\bfr_2')
      \avgl v^-(\bfr_1')v^-(\bfr_2') \avgr_V
  \right\}
  ,
\end{equation}
where the average $\avgl\cdots\avgr_V$ is with
respect to the total vortex Hamiltonian $H_V=H_V^++H_V^-$. 


%


\end{document}